\documentclass[10pt,twocolumn,letterpaper]{article}

\usepackage[pagenumbers]{wacv} 

\definecolor{wacvblue}{rgb}{0.21,0.49,0.74}
\usepackage[breaklinks,colorlinks,allcolors=wacvblue]{hyperref}

\def\wacvPaperID{715} 
\def\confName{WACV}
\def\confYear{2027}

\usepackage{pifont}
\usepackage{placeins}
\usepackage[table]{xcolor}
\newcommand{\cmark}{\textcolor{green!60!black}{\ding{51}}}
\newcommand{\xmark}{\textcolor{red!70!black}{\ding{55}}}
\usepackage{colortbl}
\definecolor{top1}{RGB}{245,152,153}
\definecolor{top2}{RGB}{253,205,154}
\definecolor{top3}{RGB}{248,244,140}
\usepackage{float} 

\newcommand{\partitle}[1]{\smallskip\noindent\textbf{#1}.}

\begin{document}

\title{MAGE: Modality-Agnostic Music Generation and Target-Source Extraction}

\author{
Muhammad Usama Saleem$^{1}$\thanks{Work done during a research internship at Google.},
Ravi Tejasvi$^{1}$,
Tianyu Xu$^{1}$,
Rajeev Nongpiur$^{1}$,
Ishan Chatterjee$^{1}$,\\
Mayur Jagdishbhai Patel$^{2}$,
Pu Wang$^{2}$\\
$^{1}$Google \qquad
$^{2}$University of North Carolina at Charlotte
}

 \makeatletter
\let\@oldmaketitle\@maketitle 
\renewcommand{\@maketitle}{
  \@oldmaketitle 
  \vspace{1em} 
  \centering
  \includegraphics[width=0.8\linewidth]{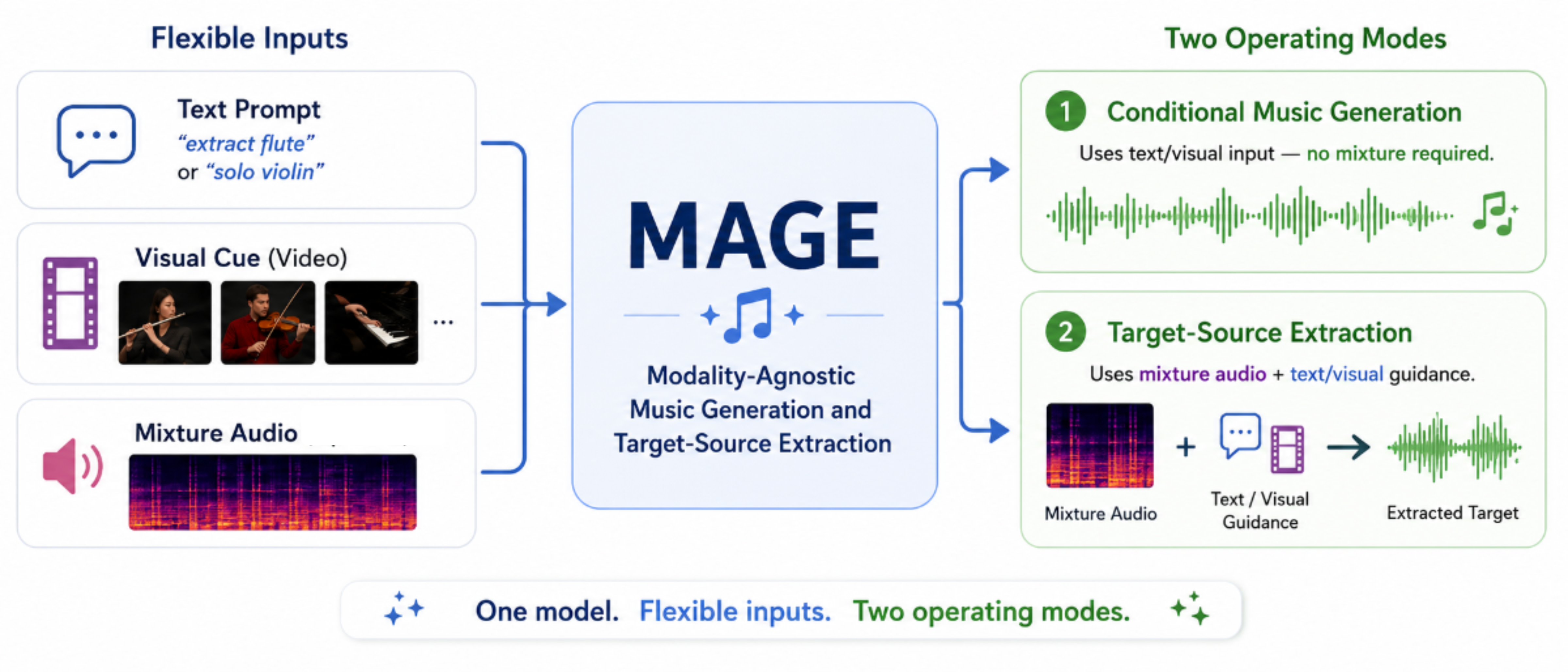}

\captionof{figure}{
\textbf{Overview of MAGE.} MAGE supports conditional music generation and mixture-grounded target-source extraction through a shared conditioning interface over text, visual, and mixture inputs. Without a reference mixture, it generates music from text, visual input, or both; with an observed mixture and a target-identifying condition, it recovers the specified source.
}

  \label{fig:landing}
  \bigskip
}
\makeatother

\maketitle

\begin{abstract}

Recent advances in multimodal audio generation have enabled music synthesis from text, visual cues, and other high-level conditions. However, most systems are designed for a single operating mode: either generating music without a reference mixture or extracting a target source from an existing mixture. This fixed-task design limits their use when different combinations of text, visual, and mixture inputs are available. To address this gap, we propose \textbf{MAGE}, a modality-agnostic framework for conditional music generation and mixture-grounded target-source extraction within a shared continuous latent space. Our approach introduces three key components. First, a \textit{Controlled Multimodal FluxFormer} models the conditional flow from noise to a target audio latent, enabling the same backbone to operate with or without a mixture condition. Second, \textit{Audio--Visual Nexus Alignment} maps frame-level visual features onto the audio latent sequence, allowing visual evidence to condition the generation process at the audio-token level. Third, a cross-gated modulation mechanism uses the aligned visual representation to regulate intermediate audio features, while text provides separate semantic guidance. We further train MAGE with dynamic modality masking, exposing the same model to text-only, visual-only, joint text--visual, mixture-conditioned, and unconditional configurations. Experiments on the MUSIC benchmark evaluate MAGE under separate protocols for mixture-free generation and mixture-grounded target-source extraction. The results show that MAGE provides a shared conditioning interface across both settings, and that the proposed alignment and gating components improve interference suppression in the extraction task.

\end{abstract}

\vspace{-20pt}
\section{Introduction}

Multimodal music systems increasingly rely on heterogeneous cues such as text, video, and audio context. Text prompts provide high-level semantic intent, such as the desired instrument or acoustic scene, while visual inputs provide evidence about visible instruments, performers, and scene context. In practical settings, however, these cues may be available under different operating conditions. A user may want to synthesize music from text or video when no reference audio exists, or recover a specified source from an observed mixture when mixture audio is available. These two settings are related but distinct: conditional music generation produces audio without a reference mixture~\cite{copet2023simple,liu2024audioldm2,prajwalmusicflow,tian2025audiox,zuo2025gvmgen}, whereas target-source extraction recovers a specified source from an observed mixture using text, visual, or other target-identifying conditions~\cite{AudioSep,dong2023clipsep,flowsep,wang2025soloaudio,huang2025davis,shi2025samaudio,wen2025promptsep}. We address these two settings within a shared conditional latent model: conditional music generation and mixture-grounded target-source extraction.

Existing methods are typically designed for one side of this problem. Music demixing and audio--visual separation models estimate target sources from mixtures and remain strong task-specific baselines~\cite{zhao2018sound,gao2019co,gan2020music,tian2021cyclic,chatterjee2021visual,zhu2022visually,chen2023iquery,defossez2019music,defossez2021hybrid,rouard2022hybrid,Lu2023MusicSS,musdb18}. Promptable extraction systems further allow the target source to be specified through language, visual prompts, or other flexible cues~\cite{AudioSep,dong2023clipsep,flowsep,wang2025soloaudio,huang2025davis,shi2025samaudio,wen2025promptsep}. These methods, however, are grounded in an observed mixture and do not address mixture-free music generation. Conversely, conditional generation models synthesize music from text, video, or other semantic conditions, but generally do not recover a specified source from an existing mixture~\cite{copet2023simple,liu2024audioldm2,prajwalmusicflow,tian2025audiox,zuo2025gvmgen}. This creates a gap between generation systems that operate without mixture audio and extraction systems that require it.

We introduce \textbf{MAGE}, a modality-agnostic framework that supports conditional music generation and mixture-grounded target-source extraction within a shared continuous latent formulation. MAGE represents audio using a learned continuous codec and uses a \textit{Controlled Multimodal FluxFormer} to model a conditional flow from noise to the target audio latent. When no mixture is provided, the model generates music from text, visual input, or both. When a mixture is available, its encoded latent is supplied as an acoustic condition, while text or visual input specifies the source to be recovered. Both operating modes use the same model parameters and flow-based inference procedure. A central challenge is that text, visual input, and mixture audio provide different types of information. Text provides global semantic guidance, visual frames provide scene-level evidence, and mixture audio provides acoustic context for extraction. Directly fusing these signals can obscure their roles and does not align frame-level visual evidence with the audio latent sequence. MAGE addresses this with \textit{Audio--Visual Nexus Alignment} (AVNA), which maps visual frame features to the audio latent resolution before conditioning. The aligned visual representation is then used by cross-gated modulation to regulate intermediate audio features, while text is handled through a separate semantic-conditioning pathway. This design allows visual information to condition the audio-token stream without treating text as temporally aligned evidence.

MAGE is trained with dynamic modality masking, exposing the same model to text-only, visual-only, joint text--visual, mixture-conditioned, and unconditional configurations. We use the term \emph{modality-agnostic} to describe the shared conditioning interface through which the model handles the supported text, visual, mixture-conditioned, and unconditional configurations. We evaluate MAGE on MUSIC using separate protocols for mixture-free generation and mixture-grounded target-source extraction~\cite{zhao2018sound}. For generation, we evaluate audio quality and conditioning alignment. For extraction, we evaluate reconstruction quality and interference suppression against specialized and promptable baselines. Dedicated separation systems are treated as valid task-specific baselines, while MAGE evaluates how a shared conditional model performs across both mixture-free generation and mixture-grounded extraction.

Our contributions are summarized as follows:
\begin{itemize}
    \item We propose \textbf{MAGE}, a modality-agnostic framework that treats conditional music generation and mixture-grounded target-source extraction as two operating modes of a shared continuous latent model.

    \item We introduce the \textit{Controlled Multimodal FluxFormer}, a flow-matching Transformer that predicts target audio latents under supported combinations of text, visual, and mixture conditions using the same model parameters and inference procedure.

    \item We propose \textit{Audio--Visual Nexus Alignment} and cross-gated visual modulation to map frame-level visual features onto the audio latent sequence, and train the model with dynamic modality masking to support text-only, visual-only, joint text--visual, mixture-conditioned, and unconditional configurations.

    \item We conduct separate evaluations for mixture-free generation and mixture-grounded target-source extraction on MUSIC, compare extraction against specialized and promptable baselines, and analyze the effects of alignment, gated visual conditioning, and modality masking through ablations.
\end{itemize}

\vspace{-10pt}
\section{Related Work}
\label{sec:related_work}
\vspace{-5pt}

\partitle{Music Demixing and Target-Source Extraction}
Music source separation aims to recover target sources from an observed mixture. Dedicated demixing systems such as Demucs~\cite{defossez2019music}, Hybrid Demucs~\cite{defossez2021hybrid,rouard2022hybrid}, and related architectures~\cite{Lu2023MusicSS} achieve strong reconstruction performance on benchmarks such as MUSDB18~\cite{musdb18}. Audio--visual methods further use visual evidence to identify the source of interest, including Sound-of-Pixels~\cite{zhao2018sound}, Co-Separation~\cite{gao2019co}, Music Gesture~\cite{gan2020music}, CCoL~\cite{tian2021cyclic}, audio--visual correspondence models~\cite{chatterjee2021visual,zhu2022visually}, iQuery~\cite{chen2023iquery}, and DAVIS~\cite{huang2025davis}. Promptable extraction methods such as Universal Sound Separation~\cite{kong2023universal}, AudioSep~\cite{AudioSep}, CLIPSep~\cite{dong2023clipsep}, FlowSep~\cite{flowsep}, SoloAudio~\cite{wang2025soloaudio}, SAM Audio~\cite{shi2025samaudio}, and PromptSep~\cite{wen2025promptsep} broaden the target specification interface through language, visual, or other prompts. These methods remain grounded in an observed mixture, whereas MAGE supports both mixture-grounded extraction and mixture-free generation within the same conditional latent model.

\partitle{Conditional and Multimodal Music Generation}
Conditional music-generation models synthesize audio from semantic or perceptual cues without requiring a reference mixture. MusicGen~\cite{copet2023simple}, AudioLDM 2~\cite{liu2024audioldm2}, and MusicFlow~\cite{prajwalmusicflow} generate music from language descriptions using autoregressive, diffusion, or flow-based modeling. Video-to-music systems such as GVMGen~\cite{zuo2025gvmgen} use visual features to improve video--music correspondence, while AudioX~\cite{tian2025audiox} supports broader multimodal audio generation through masked multimodal training. These systems focus on mixture-free generation, whereas MAGE supports both mixture-free generation and mixture-grounded target-source extraction within a shared conditional latent model and evaluates them under separate task-specific protocols.

\partitle{Multimodal Conditioning and Audio--Visual Alignment}
Multimodal audio models combine signals with different roles: text provides semantic guidance, visual frames provide scene-level evidence, and mixture audio provides acoustic context. Prior systems such as CLIPSep~\cite{dong2023clipsep}, AudioSep~\cite{AudioSep}, DAVIS~\cite{huang2025davis}, GVMGen~\cite{zuo2025gvmgen}, and AudioX~\cite{tian2025audiox} incorporate these signals through feature fusion, attention, adaptive modulation, or masked multimodal training. Audio--visual methods such as Co-Separation~\cite{gao2019co}, Music Gesture~\cite{gan2020music}, CCoL~\cite{tian2021cyclic}, iQuery~\cite{chen2023iquery}, and GVMGen~\cite{zuo2025gvmgen} further address cross-modal mismatch using object localization, motion cues, cyclic consistency, query-based interaction, or hierarchical temporal modeling. MAGE differs by mapping frame-level visual features to the audio latent resolution with Audio--Visual Nexus Alignment, applying cross-gated visual modulation, and using modality masking so the same model operates under text-only, visual-only, joint text--visual, mixture-conditioned, and unconditional configurations.


\begin{figure*}[ht]
    \centering
    \includegraphics[width=0.9\linewidth]{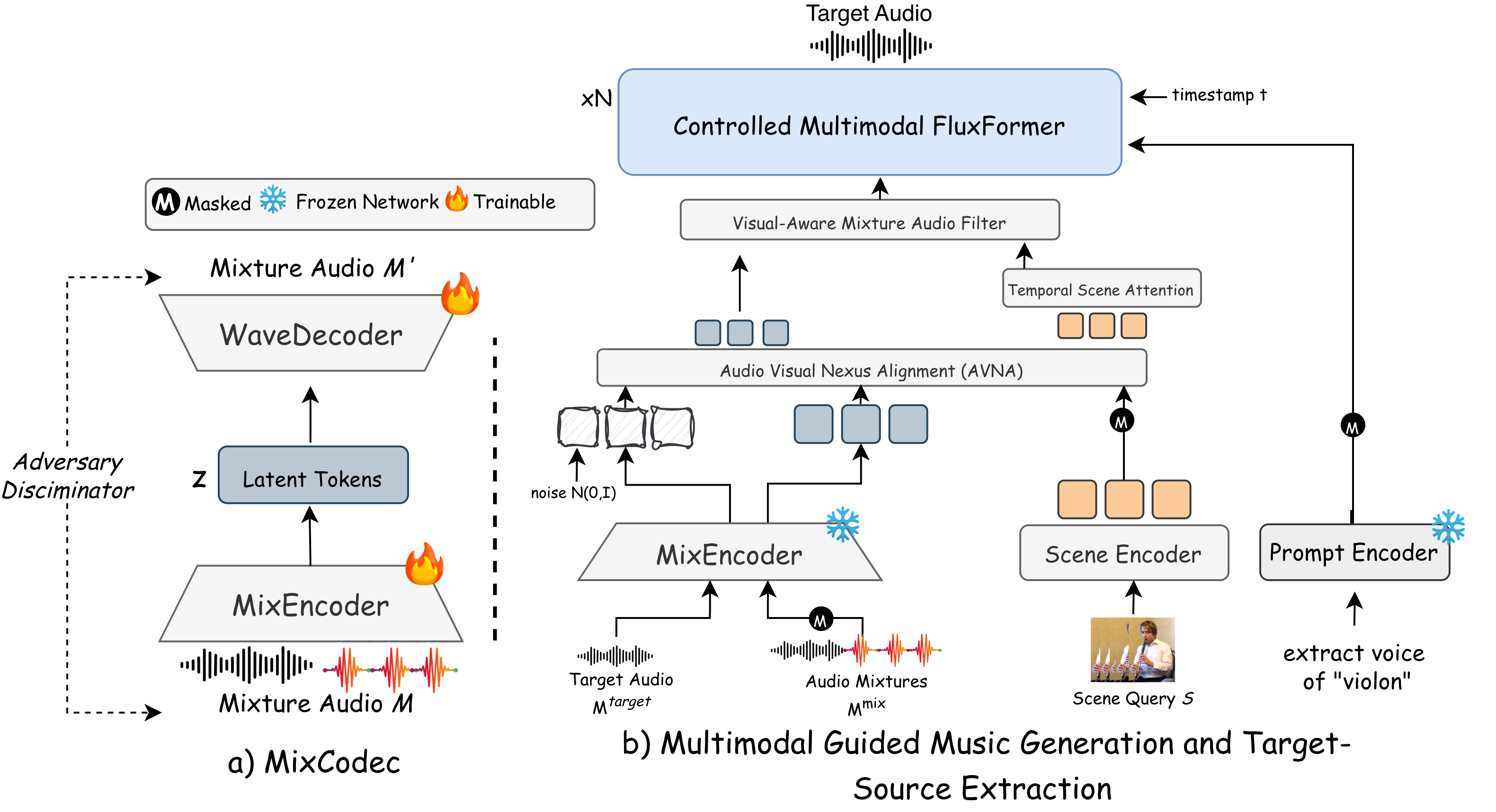}

\caption{
Overview of MAGE. MixWavCodec maps waveforms to continuous audio latents and reconstructs audio from generated latents. The Controlled Multimodal FluxFormer models a conditional flow from noise to the target audio latent under supported text, visual, and mixture conditions. AVNA maps frame-level visual features to the audio latent sequence, and cross-gated modulation applies visual conditioning while text provides semantic guidance. The same model supports mixture-free conditional generation and mixture-grounded target-source extraction.
}

    \label{fig:overview}
\end{figure*}

\vspace{-5pt}

\section{Proposed Method: MAGE}
\label{sec:method}
\vspace{-5pt}

\partitle{Problem Formulation}
We study two related conditional audio-generation tasks: \emph{conditional music generation} without a reference mixture and \emph{mixture-grounded target-source extraction}. Let $x^\star \in \mathbb{R}^{T}$ denote the target waveform and $\hat{x} \in \mathbb{R}^{T}$ denote the model output. The available inputs may include a mixture waveform $x_m$, a visual input $v$ given as a video clip or sampled frames, and a text prompt $c$ describing either the intended musical content or the target source. In the generation setting, no mixture is provided, and the model samples $\hat{x} \sim p_{\theta}(x \mid \mathcal{C}_{\mathrm{gen}})$, where $\mathcal{C}_{\mathrm{gen}} \in \{\{c\}, \{v\}, \{c,v\}\}$. In the extraction setting, the observed mixture is provided together with at least one target-identifying condition, and the model samples $\hat{x} \sim p_{\theta}(x \mid x_m, \mathcal{C}_{\mathrm{ext}})$, where $\mathcal{C}_{\mathrm{ext}} \in \{\{c\}, \{v\}, \{c,v\}\}$. The mixture supplies the acoustic context, while the text or visual condition specifies the source to be recovered; we therefore exclude mixture-only extraction, since the mixture alone does not define a unique target. We refer to the mixture-conditioned task as \emph{target-source extraction}. Our goal is to support both operating modes with a single model, a shared continuous latent representation, and a common inference procedure without task-specific output heads.

\partitle{Method Overview}
MAGE represents audio in a continuous latent space and models both conditional music generation and mixture-grounded target-source extraction with a shared flow-based backbone, as shown in Fig.~\ref{fig:overview}. An audio waveform is first mapped to latent tokens using \textit{MixWavCodec}, whose encoder and decoder are trained to reconstruct audio from the latent representation. The target latent is then generated by the \textit{Controlled Multimodal FluxFormer}, a Transformer-based flow model that learns a conditional trajectory from noise to the target audio latent. When no mixture is provided, the model generates audio from text, visual input, or both. When a mixture is available, its encoded latent is supplied as an additional acoustic condition, while text or visual input specifies the target source to be recovered. To incorporate visual information, MAGE uses \textit{Audio--Visual Nexus Alignment} (AVNA) to map frame-level visual features onto the audio latent sequence. The aligned visual representation is then used by a cross-gated modulation mechanism to condition intermediate audio features, while text is introduced through a separate semantic-conditioning pathway. During training, dynamic modality masking exposes the same model to text-only, visual-only, joint text--visual, mixture-conditioned, and unconditional configurations. This allows a single model to be evaluated across the supported generation and extraction settings without task-specific output heads.

\vspace{-5pt}

\subsection{Continuous Audio Representation}
\label{sec:mixwavcodec}

MAGE operates in a continuous audio-latent space defined by a learned waveform codec, which we refer to as \textit{MixWavCodec}. The codec reduces the temporal length of the audio sequence before flow modeling and decodes generated latents back to waveform space. Given an audio waveform $x\in\mathbb{R}^{T}$, the encoder maps it to latent tokens $z=\mathcal{E}_{\phi}(x)\in\mathbb{R}^{N\times D}$, and the decoder reconstructs the waveform as $\tilde{x}=\mathcal{D}_{\phi}(z)$, where $N$ is the number of latent tokens and $D$ is the latent dimension. We use a continuous bottleneck rather than discrete audio tokens so that the flow model can predict target audio latents directly in a differentiable latent space. The codec is trained as an audio reconstruction model using both isolated sources and randomly mixed audio segments. The mixture examples are used as data augmentation to expose the codec to overlapping musical content encountered in the extraction setting, while the main modeling contribution lies in the shared conditional latent flow built on top of this representation. For two sampled audio segments $x^{(1)}$ and $x^{(2)}$, we form a training mixture $x_m=x^{(1)}+\lambda x^{(2)}$, where $\lambda\sim\mathcal{U}[\lambda_{\min},\lambda_{\max}]$.

To make the decoder less sensitive to small deviations in latent space, we apply latent perturbation during codec training. Specifically, we perturb the encoded latent as $\tilde{z}=z+\sigma\epsilon$, where $\epsilon\sim\mathcal{N}(0,I)$, and reconstruct the waveform from $\tilde{z}$. This augmentation exposes the decoder to latent states near the encoded data manifold, which may occur when the flow model generates latents at inference time. The codec is optimized with reconstruction and adversarial objectives:
\begin{equation}
    \min_{\mathcal{E}_{\phi},\mathcal{D}_{\phi}}
    \mathcal{L}_{\mathrm{rec}}
    + \alpha \mathcal{L}_{\mathrm{adv}}
    + \beta \mathcal{L}_{\mathrm{fm}},
    \qquad
    \min_{\mathcal{D}_{\mathrm{adv}}}
    \mathcal{L}_{\mathcal{D}_{\mathrm{adv}}}.
    \label{eq:codec_objective}
\end{equation}
Here, $\mathcal{L}_{\mathrm{rec}}$ combines waveform and spectral reconstruction terms, $\mathcal{L}_{\mathrm{adv}}$ is the adversarial loss, and $\mathcal{L}_{\mathrm{fm}}$ is the feature-matching loss. The discriminator is used only for codec training. In the full MAGE pipeline, the trained codec provides the latent representation for both mixture-free generation and mixture-grounded target-source extraction. The codec and latent perturbation are evaluated through reconstruction metrics and downstream ablations to quantify their effect on the shared latent modeling pipeline.

\vspace{-5pt}

\subsection{Conditional Flow Matching}
\label{sec:flow_matching}

MAGE models target audio generation as conditional flow matching in the continuous latent space of the pretrained codec. Let $z_1=\mathcal{E}_{\phi}(x^\star)\in\mathbb{R}^{N\times D}$ denote the target audio latent, and let $z_0\sim\mathcal{N}(0,I)$ be a Gaussian noise latent of the same shape. For $t\sim\mathcal{U}(0,1)$, we construct the interpolation $z_t=(1-t)z_0+t z_1$ with target velocity $u_t=z_1-z_0$. The \textit{Controlled Multimodal FluxFormer} $F_{\theta}$ predicts the conditional velocity $\hat{u}_t=F_{\theta}(z_t,\widetilde{\mathcal{C}},t)$, where $\widetilde{\mathcal{C}}$ is the effective condition set after modality masking. The model is trained with the velocity-matching objective
\begin{equation}
    \mathcal{L}_{\mathrm{FM}}
    =
    \mathbb{E}_{z_0,z_1,t,\widetilde{\mathcal{C}}}
    \left[
        \left\|
        F_{\theta}(z_t,\widetilde{\mathcal{C}},t)-u_t
        \right\|_{1}
    \right].
    \label{eq:flow_matching_loss}
\end{equation}

In the generation setting, $\widetilde{\mathcal{C}}$ contains text, visual input, or both, with the mixture branch set to its null condition. In the extraction setting, $\widetilde{\mathcal{C}}$ additionally contains the mixture latent $z_m=\mathcal{E}_{\phi}(x_m)$ and at least one target-identifying condition. At inference, we sample $z_0\sim\mathcal{N}(0,I)$, integrate the learned vector field from $t=0$ to $t=1$, obtain the predicted target latent $\hat{z}_1$, and reconstruct the waveform as $\hat{x}=\mathcal{D}_{\phi}(\hat{z}_1)$. The same FluxFormer parameters and inference procedure are used for both mixture-free generation and mixture-grounded target-source extraction; only the selected condition set changes.

\partitle{Multimodal Condition Encoders}
MAGE conditions the latent flow on text, visual input, and, when available, mixture audio. A text encoder $\phi_{\mathrm{text}}$ maps the prompt $c$ to a semantic embedding
\begin{equation}
    p = \phi_{\mathrm{text}}(c),
    \label{eq:text_encoder}
\end{equation}
which provides global guidance about the intended musical content or the target source. A visual encoder $\phi_{\mathrm{vis}}$ extracts frame-level features from a visual input $v=\{I_k\}_{k=1}^{K}$:
\begin{equation}
    s_k = \phi_{\mathrm{vis}}(I_k),
    \qquad k=1,\ldots,K .
    \label{eq:visual_encoder}
\end{equation}
For mixture-grounded target-source extraction, the mixture waveform is encoded by the audio codec as
\begin{equation}
    z_m = \mathcal{E}_{\phi}(x_m).
    \label{eq:mixture_encoder}
\end{equation}
The mixture latent provides acoustic context, while the text or visual condition identifies the target source to be recovered. When a modality is unavailable, it is replaced by a null condition and handled through the modality-masking strategy described in Sec.~\ref{sec:masking}.

\partitle{Audio--Visual Nexus Alignment}
The audio latent sequence contains $N$ tokens, whereas the visual encoder produces $K$ frame-level features. Directly fusing these sequences is not well defined because they have different temporal resolutions. To address this, \textit{Audio--Visual Nexus Alignment} (AVNA) maps the frame-level visual features to the audio latent resolution before visual conditioning is applied.

Let $\tau_i^a$ and $\tau_k^v$ denote normalized positions of audio latent token $i$ and visual frame $k$:
\begin{equation}
    \tau_i^a = \frac{i-1}{N-1},
    \qquad
    \tau_k^v = \frac{k-1}{K-1}.
\end{equation}
For each audio latent index $i$, AVNA assigns the nearest visual feature in normalized time:
\begin{equation}
    \pi(i) = \arg\min_k \left|\tau_i^a-\tau_k^v\right|,
    \qquad
    \bar{s}_i = s_{\pi(i)} .
    \label{eq:avna}
\end{equation}
This produces an aligned visual sequence $\bar{s}\in\mathbb{R}^{N\times d_v}$ with the same length as the audio latent sequence. AVNA provides feature-resolution alignment between sampled visual frames and audio latent tokens, allowing visual cues to be applied consistently across the latent trajectory. Since the visual encoder operates on sampled frames, the aligned features primarily capture instrument and scene-level evidence rather than explicit onset-level motion cues.

\partitle{Cross-Gated Visual Modulation}
After AVNA, the aligned visual features are used to modulate intermediate audio representations in the FluxFormer. Let $h_i^{(\ell)}$ denote the audio feature at layer $\ell$ and latent index $i$. We compute a visual gate
\begin{equation}
    g_i^{(\ell)}
    =
    \sigma
    \left(
        W_s^{(\ell)}\bar{s}_i + b_s^{(\ell)}
    \right),
    \label{eq:visual_gate}
\end{equation}
where $\sigma(\cdot)$ is a sigmoid activation and $W_s^{(\ell)}$ is a learnable projection. The gated audio feature is then
\begin{equation}
    \tilde{h}_i^{(\ell)}
    =
    h_i^{(\ell)}
    \odot
    g_i^{(\ell)} .
    \label{eq:cgm}
\end{equation}

This modulation lets visual information guide the audio-token representation without directly injecting visual features into the audio stream. Text is handled through a separate semantic-conditioning pathway, reflecting its role as global guidance rather than frame-level evidence. Together, AVNA and cross-gated modulation provide a structured mechanism for incorporating visual cues into the latent flow. Their contribution is evaluated through ablations on generation quality, extraction quality, and interference suppression.

\subsection{Dynamic Modality Masking and Inference}
\label{sec:masking}

\partitle{Dynamic Modality Masking}
MAGE is trained with dynamic modality masking so the same conditional flow model can support multiple valid conditioning settings. Let the available condition set be $\mathcal{C}=\{z_m,p,\bar{s}\}$, where $z_m=\mathcal{E}_{\phi}(x_m)$ is the mixture latent, $p=\phi_{\mathrm{text}}(c)$ is the text embedding, and $\bar{s}$ is the AVNA-aligned visual sequence. During training, we sample a binary mask $m=(m_m,m_t,m_v)\in\{0,1\}^{3}$ and construct the effective condition set $\widetilde{\mathcal{C}}=\{\tilde{z}_m,\tilde{p},\tilde{s}\}$, where missing modalities are replaced by null conditions: $\tilde{z}_m=m_m z_m+(1-m_m)z_{\varnothing}$, $\tilde{p}=m_t p+(1-m_t)p_{\varnothing}$, and $\tilde{s}=m_v\bar{s}+(1-m_v)s_{\varnothing}$.

We sample masks only from supported task configurations. For mixture-free generation, the valid masks are $\mathcal{M}_{\mathrm{gen}}=\{(0,1,0),(0,0,1),(0,1,1)\}$, corresponding to text-only, visual-only, and joint text--visual conditioning. For mixture-grounded target-source extraction, the valid masks are $\mathcal{M}_{\mathrm{ext}}=\{(1,1,0),(1,0,1),(1,1,1)\}$, where the mixture is paired with at least one target-identifying condition. We exclude the mixture-only mask $(1,0,0)$ because the mixture alone does not specify which source should be recovered. The null mask $(0,0,0)$ is also sampled during training for unconditional/CFG learning.

Given target latent $z_1=\mathcal{E}_{\phi}(x^\star)$, noise $z_0\sim\mathcal{N}(0,I)$, and $t\sim\mathcal{U}(0,1)$, we form $z_t=(1-t)z_0+t z_1$ with target velocity $u_t=z_1-z_0$. The FluxFormer predicts $\hat{u}_t=F_{\theta}(z_t,\widetilde{\mathcal{C}},t)$ and is trained with the masked flow-matching loss
\begin{equation}
    \mathcal{L}_{\mathrm{FM}}
    =
    \mathbb{E}_{z_0,z_1,t,m}
    \left[
        \left\|
        F_{\theta}(z_t,\widetilde{\mathcal{C}},t)-u_t
        \right\|_1
    \right].
    \label{eq:masked_fm_loss}
\end{equation}
This trains one conditional vector field across the supported generation and extraction settings without task-specific output heads.

\partitle{Inference}
At inference time, no random masking is applied; the condition set is chosen according to the requested operating mode. For mixture-free generation, the mixture branch is set to its null condition and the model is conditioned on text, visual input, or both. For mixture-grounded target-source extraction, the mixture latent $z_m=\mathcal{E}_{\phi}(x_m)$ is provided together with at least one target-identifying condition. Starting from $z_0\sim\mathcal{N}(0,I)$, we integrate the learned vector field from $t=0$ to $t=1$ to obtain $\hat{z}_1$ and decode the waveform as $\hat{x}=\mathcal{D}_{\phi}(\hat{z}_1)$. The same model parameters, ODE solver, and decoder are used for all supported configurations; only the selected condition set changes.

\partitle{Classifier-Free Guidance}
We optionally use classifier-free guidance (CFG) at inference. During training, the null-conditioning mask $(0,0,0)$ is sampled with non-zero probability, allowing the FluxFormer to learn both conditional and unconditional vector fields. Given a valid condition set $\widetilde{\mathcal{C}}$, we compute $\hat{u}_t^{\mathrm{cond}}=F_{\theta}(z_t,\widetilde{\mathcal{C}},t)$ and $\hat{u}_t^{\mathrm{uncond}}=F_{\theta}(z_t,\mathcal{C}_{\varnothing},t)$, where $\mathcal{C}_{\varnothing}$ denotes the null condition for all modalities. The guided velocity is
\begin{equation}
    \hat{u}_t^{\mathrm{cfg}}
    =
    \hat{u}_t^{\mathrm{uncond}}
    +
    \gamma
    \left(
        \hat{u}_t^{\mathrm{cond}}
        -
        \hat{u}_t^{\mathrm{uncond}}
    \right),
    \label{eq:cfg}
\end{equation}
where $\gamma$ is the guidance scale. We keep $\gamma$ fixed for reported comparisons unless otherwise stated.


\begin{figure}[t]
    \centering
    \includegraphics[width=\linewidth]{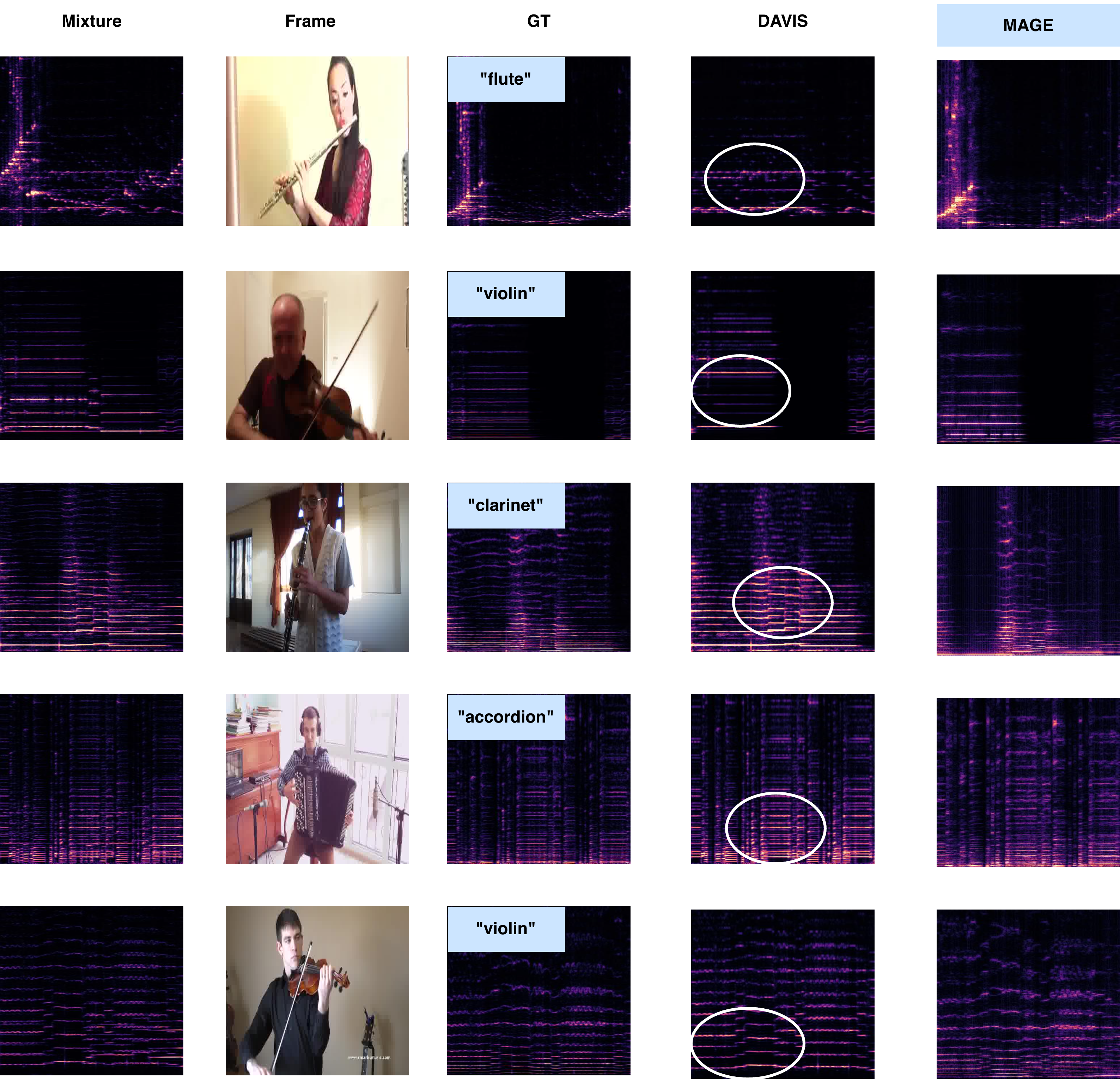}
    \caption{
    Qualitative spectrogram examples for mixture-grounded target-source extraction. Each example shows the input mixture, reference visual frame, ground-truth target spectrogram, and MAGE prediction. Highlighted regions are shown for visual inspection; quantitative extraction results are reported in Table~\ref{tab:music_extraction_main}.
    }
    \label{fig:spectrogram_examples}
    \vspace{-10pt}
\end{figure}

\begin{table*}[t]
\caption{
Mixture-grounded target-source extraction results on MUSIC~\cite{zhao2018sound}. 
SDR, SIR, and SAR evaluate extraction quality only. 
The final column indicates whether the same method also supports zero-mixture music generation, and is included only to summarize task coverage rather than to rank extraction performance. 
Best extraction results are shown in \textbf{bold} and second-best results are \underline{underlined}.
}
\label{tab:music_extraction_main}
\centering
\scalebox{0.92}{
\begin{tabular}{l|l|ccc|c}
\toprule
Method 
& Venue 
& SDR ($\uparrow$) 
& SIR ($\uparrow$) 
& SAR ($\uparrow$) 
& Zero-Mixture Generation \\
\midrule

NMF-MFCC~\cite{spiertz2009source} 
& DAFx'09 
& 0.92 & 5.68 & 6.84 
& {\color{red!70!black}\xmark} \\

Sound-of-Pixels~\cite{zhao2018sound} 
& ECCV'18 
& 4.23 & 9.39 & 9.85 
& {\color{red!70!black}\xmark} \\

Co-Separation~\cite{gao2019co} 
& ICCV'19 
& 6.54 & 11.37 & 9.46 
& {\color{red!70!black}\xmark} \\

CCoL~\cite{tian2021cyclic} 
& CVPR'21 
& 7.74 & 13.22 & 11.54 
& {\color{red!70!black}\xmark} \\

iQuery~\cite{chen2023iquery} 
& CVPR'23 
& \underline{11.17} & 15.84 & \underline{14.27} 
& {\color{red!70!black}\xmark} \\

DAVIS~\cite{huang2024accv} 
& ACCV'24 
& \textbf{11.61} & \textbf{18.36} & \textbf{14.70} 
& {\color{red!70!black}\xmark} \\

\midrule

\rowcolor{gray!8}
\textbf{MAGE (ours)} 
& Ours 
& 6.67 & \underline{17.99} & 7.88 
& {\color{green!50!black}\cmark} \\

\bottomrule
\end{tabular}}
\end{table*}


\begin{table}[t]

\caption{
Multimodal conditioning results for MAGE under mixture-grounded target-source extraction and mixture-free conditional generation. Text and visual columns indicate the provided conditions. CLAP text--audio and audio--audio similarities measure prompt alignment and semantic similarity to the reference target audio, respectively. Extraction metrics are reported separately in Table~\ref{tab:music_extraction_main}; $^{*}$ denotes the ground-truth audio--audio upper bound.
}

\label{tab:multimodal_conditioning}
\centering
\scalebox{0.65}{
\begin{tabular}{lcc|cc}
\toprule
\textbf{Operating Mode / Conditioning} 
& \textbf{Text} 
& \textbf{Visual} 
& \multicolumn{2}{c}{\textbf{CLAP Similarity} $\uparrow$} \\
\cmidrule(lr){4-5}
& & 
& \textbf{Text--Audio} 
& \textbf{Audio--Audio} \\
\midrule

\multicolumn{5}{c}{\textbf{Reference}} \\
Ground truth target audio  
& \cmark & \cmark 
& 0.302 
& 1.000$^{*}$ \\

\midrule
\multicolumn{5}{c}{\textbf{Mixture-Grounded Target-Source Extraction}} \\

Text-only condition   
& \cmark & \xmark 
& 0.205 
& 0.711 \\

Visual-only condition 
& \xmark & \cmark 
& 0.203 
& 0.713 \\

\rowcolor{gray!8}
Text + visual condition
& \cmark & \cmark 
& \textbf{0.216} 
& \textbf{0.768} \\

\midrule
\multicolumn{5}{c}{\textbf{Mixture-Free Conditional Generation}} \\

Text-only condition   
& \cmark & \xmark 
& 0.194 
& 0.562 \\

Visual-only condition 
& \xmark & \cmark 
& \textbf{0.195} 
& \textbf{0.566} \\

Text + visual condition 
& \cmark & \cmark 
& 0.193 
& 0.561 \\

\bottomrule
\end{tabular}
}
\end{table}

\begin{table}[t]
\caption{
Component ablation for MAGE on mixture-grounded target-source extraction. 
Results are reported on MUSIC; higher is better.
}
\label{tab:component_ablation}
\centering
\scalebox{0.92}{
\begin{tabular}{lccc}
\toprule
\textbf{Configuration} 
& \textbf{SDR} 
& \textbf{SIR} 
& \textbf{SAR} \\
\midrule
w/o AVNA 
& 3.01 & 5.83 & 6.24 \\
w/o CGM  
& 3.23 & 5.64 & 5.32 \\
\midrule
\rowcolor{gray!8}
\textbf{MAGE (full)} 
& \textbf{6.67} 
& \textbf{17.99} 
& \textbf{7.88} \\
\bottomrule
\end{tabular}
}
\end{table}

\begin{table}[t]
\caption{
Fusion operator ablation for MAGE on mixture-grounded target-source extraction. 
All variants use the same training setup; higher is better.
}
\label{tab:fusion_ablation}
\centering
\scalebox{0.92}{
\begin{tabular}{lccc}
\toprule
\textbf{Fusion Operator} 
& \textbf{SDR} 
& \textbf{SIR} 
& \textbf{SAR} \\
\midrule
Additive fusion  
& 2.09 & 3.53 & 4.32 \\
Self-attention   
& 2.42 & 4.64 & 5.45 \\
Cross-attention  
& 2.49 & 5.64 & 6.45 \\
\midrule
\rowcolor{gray!8}
\textbf{Cross-gated modulation} 
& \textbf{6.67} 
& \textbf{17.99} 
& \textbf{7.88} \\
\bottomrule
\end{tabular}
}
\end{table}

\vspace{-10pt}

\section{Experiments}

\label{subsec:exp_setup}

\partitle{Dataset}
We evaluate MAGE on the MUSIC dataset~\cite{zhao2018sound}, a widely used benchmark for audio--visual musical source separation. MUSIC contains in-the-wild videos of musical performances with instrument-level labels and corresponding audio tracks. We use the dataset for two evaluation settings. For mixture-grounded target-source extraction, the model receives an observed mixture together with text and/or visual conditions that identify the target source. For mixture-free generation, the model receives text, visual input, or both, but no mixture audio. Instrument labels are converted into simple text prompts, while sampled video frames provide the visual condition. We follow the standard train/validation/test protocol used by prior MUSIC evaluations and keep the generation and extraction test settings separate.

\partitle{Evaluation Protocols and Metrics}
We evaluate MAGE under two task-specific protocols. For \emph{mixture-grounded target-source extraction}, we report standard source-separation metrics: SDR, SIR, and SAR. SDR measures overall reconstruction quality, SIR measures suppression of interfering sources, and SAR measures artifact-related distortion. These metrics are computed only when a ground-truth target source is available.

For \emph{mixture-free generation}, separation metrics are not applicable because there is no observed mixture and no uniquely defined separated target. We therefore report generation-oriented metrics that measure audio realism and conditioning alignment. Specifically, we use CLAP-based audio--audio similarity to measure semantic similarity between generated and reference audio, and CLAPScore to evaluate alignment between generated audio and the text condition. Because text conditioning and automatic evaluation both use CLAP-based representations, these metrics may favor semantic alignment captured by CLAP. We therefore interpret them as automatic alignment indicators rather than substitutes for perceptual listening evaluation.

This separation of protocols is important because MAGE supports both operating modes, but the two modes measure different capabilities. Extraction evaluates recovery of a specified source from a mixture, while generation evaluates whether the model can synthesize plausible and condition-aligned music without using mixture audio.

\partitle{Notation}
Let $x \in \mathbb{R}^{T}$ denote an audio waveform, $x_m$ denote an observed mixture waveform when available, and $x^\star$ denote the ground-truth target waveform. The codec encoder $\mathcal{E}_{\phi}$ maps audio to a continuous latent sequence, and the decoder $\mathcal{D}_{\phi}$ maps latents back to waveform space:
\begin{equation}
    z = \mathcal{E}_{\phi}(x) \in \mathbb{R}^{N \times D},
    \qquad
    \hat{x} = \mathcal{D}_{\phi}(z).
\end{equation}
We denote the target latent as $z_1=\mathcal{E}_{\phi}(x^\star)$ and, when a mixture is provided, the mixture latent as $z_m=\mathcal{E}_{\phi}(x_m)$. A text prompt $c$ is encoded as $p=\phi_{\mathrm{text}}(c)$, and a visual input $v=\{I_k\}_{k=1}^{K}$ is encoded into frame-level features $s_k=\phi_{\mathrm{vis}}(I_k)$. AVNA maps these frame-level features to the audio latent resolution, producing aligned visual tokens $\bar{s}\in\mathbb{R}^{N\times d_v}$.

To support multiple conditioning settings, we define a modality mask $m=(m_m,m_t,m_v)\in\{0,1\}^{3}$ over mixture, text, and visual conditions. The effective condition set is
\begin{equation}
    \widetilde{\mathcal{C}}
    =
    \{\tilde{z}_m,\tilde{p},\tilde{s}\},
\end{equation}
where unavailable modalities are replaced by their corresponding null conditions. We use only valid task masks: text-only, visual-only, and text--visual masks for mixture-free generation; and mixture plus at least one target-identifying condition for target-source extraction. The mixture-only configuration is excluded because the mixture alone does not specify which source should be recovered.

\partitle{Implementation Details}
We implement MAGE in PyTorch and train it on MUSIC using 64{,}640-sample audio segments at 11.025\,kHz. MixWavCodec maps each waveform to $N=202$ continuous latent tokens with dimension $D=128$. After codec training, its encoder and decoder are frozen while training the FluxFormer. For visual conditioning, we sample 47 frames at 8\,fps and encode them with CLIP ViT-B/32; AVNA then maps the frame features to the audio latent resolution. Text prompts are encoded using a frozen CLAP text encoder, with MUSIC instrument labels converted into simple text prompts. The Controlled Multimodal FluxFormer is a flow-matching Transformer with hidden size 1024 operating on the 202-token audio latent sequence. The input latent $z_t$ is projected to the model dimension and combined with learned positional embeddings.  When available, the mixture latent $z_m$ is encoded by the same codec and provided as an additional condition. Text, visual, and mixture inputs are incorporated through their respective conditioning pathways, and cross-gated modulation is applied only to AVNA-aligned visual features. 

We train the FluxFormer with the masked flow-matching objective in Eq.~\ref{eq:masked_fm_loss}, using an $\ell_1$ velocity loss in latent space. Task configurations are sampled as mixture-grounded extraction with probability 0.70, mixture-free generation with probability 0.20, and null-conditioning for unconditional/CFG training with probability 0.10. Extraction always includes the mixture plus at least one target-identifying condition; mixture-only extraction is not used. When visual conditioning is present, static-frame augmentation is applied with probability 0.25. We use AdamW with learning rate $10^{-4}$, batch size 32, gradient clipping at 5.0, and train for 2000 epochs. During this stage, the FluxFormer and trainable visual-conditioning layers are optimized, while the audio codec and CLAP text encoder remain frozen. At inference, we sample Gaussian noise in latent space and integrate the learned vector field with an Euler solver; all reported comparisons use the same integration steps and CFG scale unless otherwise stated. Additional implementation details are provided in the \textbf{Supplementary Material}.

\subsection{Comparison with Prior Methods}
\label{subsec:mage_comparison}

We evaluate MAGE under separate extraction and generation protocols. Extraction measures target-source recovery from a mixture, while generation measures synthesis quality and conditioning alignment without mixture audio.

\partitle{Mixture-Grounded Target-Source Extraction}
Table~\ref{tab:music_extraction_main} compares MAGE with representative audio--visual source-separation and target-source extraction methods on MUSIC~\cite{zhao2018sound}, including Sound-of-Pixels~\cite{zhao2018sound}, Co-Separation~\cite{gao2019co}, CCoL~\cite{tian2021cyclic}, iQuery~\cite{chen2023iquery}, and DAVIS~\cite{huang2025davis}. These methods are specialized for mixture-conditioned extraction and remain strong task-specific baselines. MAGE targets a broader operating regime, using one shared conditional model for both extraction and mixture-free generation. In the extraction setting, MAGE achieves competitive interference suppression, as reflected by SIR, while SDR and SAR reflect the reconstruction-quality advantage of specialized extraction systems.

\partitle{Mixture-Free Conditional Generation}
Table~\ref{tab:multimodal_conditioning} evaluates MAGE in the zero-mixture setting, where the model receives text, visual input, or both, but no mixture audio. This setting is not applicable to conventional separation systems because they require an observed mixture as input. We therefore evaluate mixture-free generation using CLAP-based semantic similarity and conditioning-alignment metrics. The text-only, visual-only, and text--visual results are close, indicating that MUSIC often provides overlapping instrument-level cues across modalities. This suggests that zero-mixture generation on this benchmark is primarily driven by coarse instrument semantics, while joint text--visual conditioning becomes more informative in the mixture-grounded setting where the target source must be identified within an observed mixture. These results demonstrate the additional zero-mixture operating mode supported by the same model, while extraction performance is evaluated separately with SDR, SIR, and SAR.

\vspace{-10pt}
\section{Ablation Study}
\label{sec:ablation}

We analyze the effect of MAGE's main design choices on MUSIC~\cite{zhao2018sound}. Unless otherwise stated, all ablated models use the same backbone, training schedule, conditioning encoders, and evaluation protocol. We report SDR, SIR, and SAR for mixture-grounded target-source extraction.

\partitle{Effect of AVNA and Cross-Gated Modulation}
Table~\ref{tab:component_ablation} shows that removing either Audio--Visual Nexus Alignment (AVNA) or cross-gated modulation reduces extraction performance. The drop in SIR is especially large, indicating that these components improve interference suppression in the mixture-grounded extraction setting. AVNA aligns visual features to the audio latent sequence, while cross-gated modulation uses the aligned visual features to regulate intermediate audio representations.

\partitle{Fusion Operator Comparison}
Table~\ref{tab:fusion_ablation} compares cross-gated modulation with alternative fusion operators under the same training setup. Additive fusion, self-attention, and cross-attention all yield lower SDR, SIR, and SAR than cross-gated modulation. This suggests that using aligned visual features as multiplicative gates is more effective than directly adding or attending to visual features in this setting.

\vspace{-10pt}
\section{Conclusion}
\vspace{-5pt}

We presented \textbf{MAGE}, a modality-agnostic framework for conditional music generation and mixture-grounded target-source extraction within a shared continuous latent formulation. MAGE combines a learned audio codec, a conditional flow-matching Transformer, Audio--Visual Nexus Alignment, cross-gated visual modulation, and dynamic modality masking to support multiple text, visual, and mixture-conditioning settings with the same model. Experiments on MUSIC show that MAGE operates in both mixture-free generation and mixture-grounded extraction regimes, while ablations indicate that visual alignment and gated modulation improve extraction performance, particularly interference suppression. The results show that MAGE provides a unified conditioning interface across generation and extraction, while specialized separators continue to define strong task-specific references for reconstruction quality.

\clearpage
\onecolumn

\setcounter{table}{0}
\setcounter{figure}{0}

\appendix
\section{Supplementary Material}

\renewcommand{\thesubsection}{\Alph{subsection}}
\subsection{Implementation Details}
\label{sec:implementation_details}

\partitle{Dataset Construction and Training Protocol} We construct training examples from MUSIC videos with instrument-level labels. Each example uses a fixed 5.86-second audio--video segment. During training, temporal crops are sampled from the available clip duration; during evaluation, fixed deterministic crops are used. Video frames are sampled uniformly from the same temporal window as the audio crop to keep the visual and audio inputs paired at the segment level. For mixture-grounded target-source extraction, each example is constructed from two source clips. One clip is designated as the target source and the other as the non-target source. After audio loading and augmentation, the observed mixture is formed by combining the target and non-target waveforms. The target source, non-target source, and mixture are then encoded into the codec latent space. The model receives the mixture together with at least one target-identifying condition, while the supervision target is the isolated target-source waveform. Instrument labels are converted into simple text prompts using the corresponding source label. For mixture-free conditional generation, the mixture branch is replaced by its learned null condition, and the model is conditioned on text, visual input, or both. This setting uses the same target-source audio as supervision but does not provide the observed mixture as input. During audio loading, waveform amplitudes are randomly scaled by a factor sampled from $[0.5,1.5]$ and clipped to $[-1,1]$. Standard spatial augmentations are applied to sampled video frames during training, while evaluation uses fixed resizing and center cropping. All compared extraction models are evaluated on the same held-out mixtures and target sources.

\begin{figure}[t]
\centering
\includegraphics[width=1\linewidth]{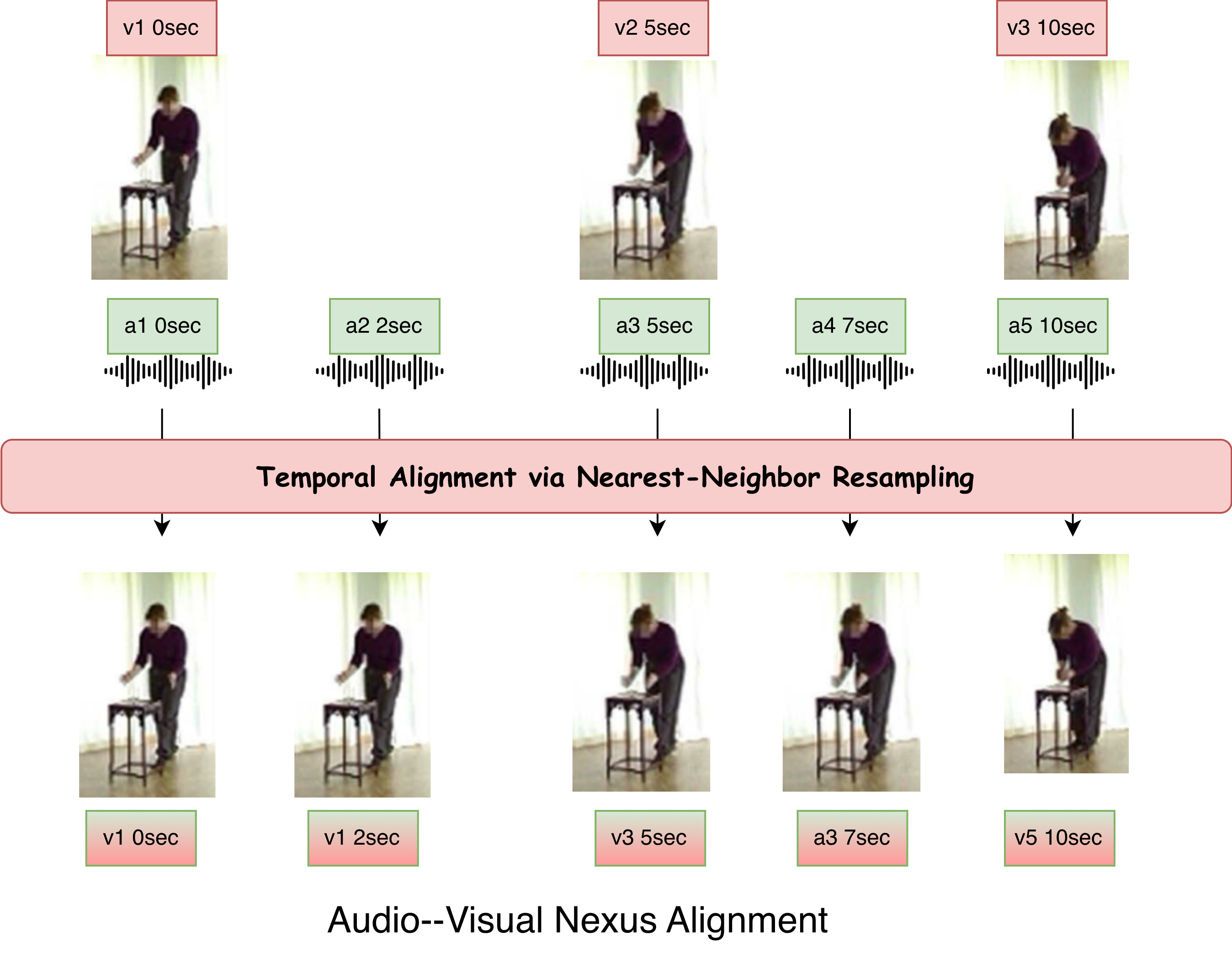}
\caption{
Illustration of \emph{Audio--Visual Nexus Alignment} (AVNA). The visual encoder produces frame-level features, while the audio codec produces a longer sequence of continuous audio latent tokens. AVNA maps the visual features to the audio latent resolution by nearest-neighbor resampling over normalized timestamps, assigning each audio latent token the closest sampled visual feature. This produces an aligned visual sequence with the same length as the audio latent sequence, which is then used for cross-gated visual modulation. AVNA provides feature-resolution alignment between sampled visual frames and audio latent tokens}
\label{fig:avna}
\end{figure}

\partitle{Audio Codec} All audio is resampled to 11.025\,kHz, converted to mono, and cropped or padded to 64{,}640 samples, corresponding to approximately 5.86 seconds. We first train \textit{MixWavCodec} as a continuous waveform codec and then use its frozen encoder and decoder as the audio representation layer for \textsc{MAGE}. The codec provides a compact continuous latent space in which the conditional flow model operates. Mixture examples are used during codec training as reconstruction augmentation, exposing the codec to overlapping musical content, but the codec itself is not treated as a standalone source-separation or generation model. The codec is trained on both isolated sources and randomly constructed mixtures so that the decoder observes the types of isolated and overlapping audio content encountered by the downstream \textsc{MAGE} pipeline. The same frozen codec is used to encode target sources, non-target sources, and observed mixtures. During \textsc{MAGE} training, the codec parameters remain fixed; the trainable components are the FluxFormer and the visual-conditioning layers.

The encoder--decoder follows an Oobleck-style architecture with weight-normalized 1D convolutions, residual units, and Snake activations. The encoder uses base width 128, channel multipliers $[1,2,4,8]$, stride pattern $[2,4,4,10]$, and a continuous VAE bottleneck. The decoder mirrors this hierarchy with transposed convolutions and residual refinement. The total temporal downsampling ratio is 320, yielding approximately 35 latent tokens per second. For each 64{,}640-sample input segment, the codec produces a continuous latent sequence of length $N=202$ with token dimension $D=128$. During codec training, we optimize waveform reconstruction, spectral reconstruction, adversarial, and feature-matching losses. We also apply latent perturbation by adding small Gaussian noise to encoded latents before decoding. This exposes the decoder to small deviations around the encoded latent manifold, which can occur when the flow model generates latent trajectories at inference time. This two-stage design separates waveform reconstruction from conditional latent modeling: MixWavCodec provides the frozen audio latent representation, while \textsc{MAGE} learns a conditional latent vector field under text, visual, and mixture conditions.

\begin{figure}[t]
\centering
\includegraphics[width=1\linewidth]{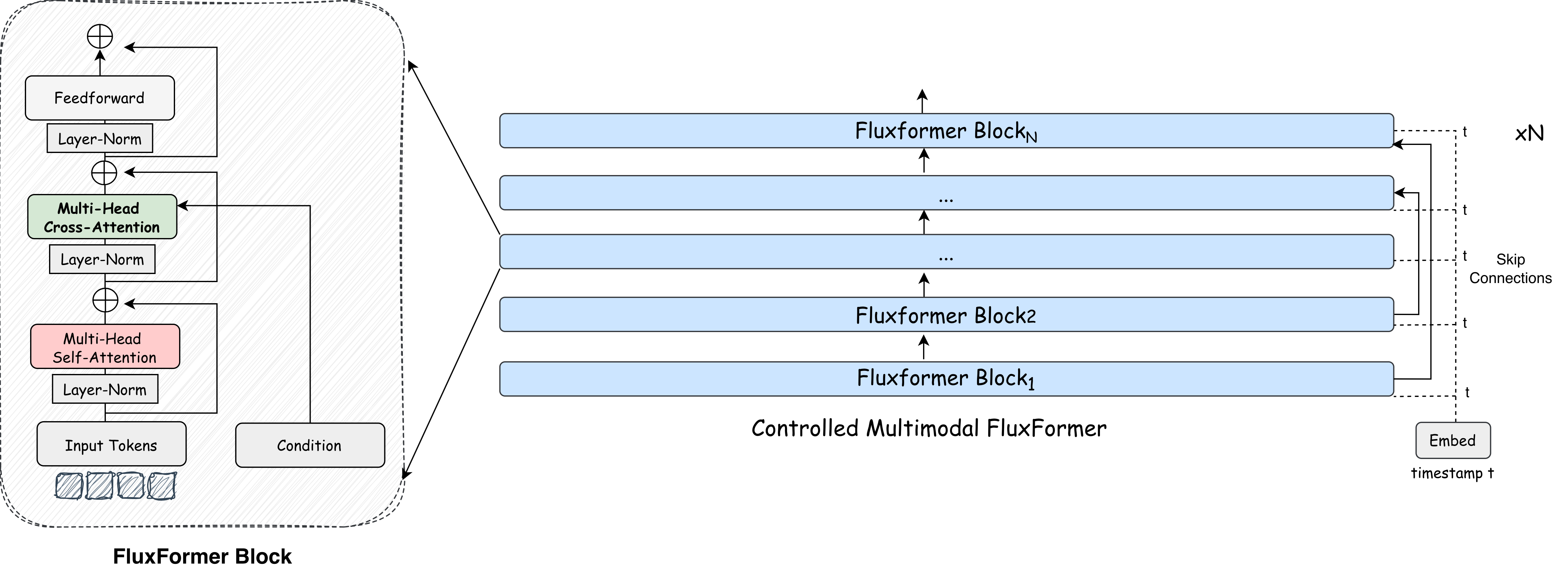}
\caption{
Architecture of the \emph{Controlled Multimodal FluxFormer}. The model operates on continuous audio latent tokens and predicts the conditional velocity field for flow matching. Text is represented by a global semantic-conditioning token, while visual features are first mapped to the audio latent resolution by AVNA and then applied through cross-gated visual modulation. When a mixture is available, its codec latent is provided as an acoustic condition; otherwise, the mixture branch is replaced by a learned null condition. The same backbone, decoder, and inference procedure are used for mixture-free conditional generation and mixture-grounded target-source extraction.
}
\label{fig:fluxformer}
\end{figure}

\partitle{Controlled Multimodal FluxFormer} As shown in Fig.~\ref{fig:fluxformer}, the generative backbone of \textsc{MAGE} is the \textit{Controlled Multimodal FluxFormer}, a Transformer-based flow model operating on continuous audio latent tokens. It uses hidden size 1024 and an encoder--bottleneck--decoder layout with skip connections between corresponding blocks. The model predicts the conditional velocity field used for flow matching in the codec latent space. At each flow step, the current latent state $z_t$ is projected from $\mathbb{R}^{128}$ to the Transformer hidden space and combined with learned positional embeddings. When a mixture is available, the mixture waveform is encoded by the same frozen audio codec to obtain $z_m$, which is projected separately and provided as an acoustic condition. The projected current latent and projected mixture latent are concatenated before being passed to the Transformer. When no mixture is provided, the mixture branch is replaced by a learned null condition. Therefore, mixture-free conditional generation and mixture-grounded target-source extraction use the same backbone, latent representation, and inference procedure; only the selected condition set changes.

\partitle{Text and Visual Conditioning} Text prompts are encoded using a frozen CLAP text encoder (\texttt{laion/clap-htsat-unfused})~\cite{wu2023clap}. We use the pooled 768-dimensional text representation, project it to the Transformer hidden space, and insert it as a global semantic-conditioning token. This provides prompt-level semantic guidance without assuming temporal correspondence between text and audio tokens. Visual conditioning is extracted using CLIP ViT-B/32~\cite{Radford2021Learning}. For each 5.86-second clip, we uniformly sample 47 frames at 8\,fps and project each frame embedding to the Transformer hidden space. The resulting frame-level visual sequence is passed to AVNA, which maps visual features to the audio latent resolution before cross-gated visual modulation. These features provide instrument- and scene-level evidence about the depicted source. Since the visual encoder operates on sampled frames rather than dense motion features, fine-grained sound-producing actions and precise audio--visual synchronization are not explicitly modeled.

\partitle{Audio--Visual Nexus Alignment} As shown in Fig.~\ref{fig:avna}, \textit{Audio--Visual Nexus Alignment} (AVNA) maps frame-level visual features to the audio latent resolution before visual conditioning is applied. The visual encoder produces $T_v$ sampled-frame features, while the audio codec produces $N$ continuous audio latent tokens. Since these sequences have different resolutions, direct token-wise fusion is not well defined. AVNA provides a deterministic feature-resolution alignment by assigning each audio latent token to the nearest sampled visual feature in normalized time.

Let $\tau_i^a=(i-1)/(N-1)$ denote the normalized position of audio latent token $i$, and let $\tau_k^v=(k-1)/(T_v-1)$ denote the normalized position of visual frame $k$. AVNA assigns
\begin{equation}
    \pi(i) = \arg\min_k |\tau_i^a-\tau_k^v|,
    \qquad
    \bar{s}_i = s_{\pi(i)} ,
\end{equation}
producing an aligned visual sequence $\bar{s}\in\mathbb{R}^{N\times d_v}$. This resampling step introduces no additional trainable alignment parameters and simply provides visual features at the same token resolution as the audio latent sequence. We therefore treat AVNA as feature-resolution alignment for conditioning.

\partitle{Cross-Gated Visual Modulation} After AVNA maps visual features to the audio latent resolution, we apply \textit{Cross-Gated Modulation} (CGM) to condition intermediate audio representations. Let $\mathbf{h}_{\mathrm{audio}}\in\mathbb{R}^{N\times d}$ denote the audio token stream at a FluxFormer layer, and let $\bar{\mathbf{s}}\in\mathbb{R}^{N\times d_v}$ denote the AVNA-aligned visual features. CGM computes a visual gate and applies it multiplicatively:
\begin{equation}
    \mathbf{g}
    =
    \sigma
    \left(
        W_s \bar{\mathbf{s}} + b_s
    \right),
    \qquad
    \mathbf{h}_{\mathrm{joint}}
    =
    \mathbf{h}_{\mathrm{audio}}
    \odot
    \mathbf{g},
    \label{eq:cgm_impl}
\end{equation}
where $W_s$ and $b_s$ are learnable parameters and $\sigma(\cdot)$ is the sigmoid function. This operation lets aligned visual features regulate the audio-token representation through a learned gate rather than through direct feature addition. Text conditioning is handled separately through a global semantic token, while visual conditioning is applied at the audio-token resolution through the aligned visual sequence. We compare CGM with additive fusion, gated residual fusion, self-attention, and cross-attention in Sec.~\ref{sec:fusion_ablation}.

\partitle{Flow-Matching Objective}
We train \textsc{MAGE} with continuous-time flow matching in the codec latent space. Let $z_1=\mathcal{E}_{\phi}(x^\star)$ denote the target audio latent, and let $z_0\sim\mathcal{N}(0,I)$ be a Gaussian noise latent with the same shape. For $t\sim\mathcal{U}(0,1)$, we define the linear interpolation path
\begin{equation}
    z_t = (1-t)z_0 + tz_1 .
    \label{eq:fm_path}
\end{equation}
The corresponding target velocity is
\begin{equation}
    u_t = z_1 - z_0 .
    \label{eq:fm_target_velocity}
\end{equation}
Given the masked condition set $\widetilde{\mathcal{C}}$, the FluxFormer predicts
\begin{equation}
    \hat{u}_t =
    F_{\theta}(z_t,\widetilde{\mathcal{C}},t).
    \label{eq:fm_prediction}
\end{equation}
We optimize the latent velocity-matching objective
\begin{equation}
    \mathcal{L}_{\mathrm{FM}}
    =
    \mathbb{E}_{z_0,z_1,t,\widetilde{\mathcal{C}}}
    \left[
    \left\|
        \hat{u}_t-u_t
    \right\|_1
    \right].
    \label{eq:fm_loss_final}
\end{equation}
This objective trains the model to predict the conditional vector field from Gaussian noise to the target audio latent. The same objective is used for all supported conditioning configurations. In mixture-free conditional generation, $\widetilde{\mathcal{C}}$ contains text, visual input, or both, with the mixture branch set to its learned null condition. In mixture-grounded target-source extraction, $\widetilde{\mathcal{C}}$ additionally includes the mixture latent and at least one target-identifying condition. Thus, the same model learns a conditional latent vector field across both operating modes; only the selected condition set changes.

\partitle{Dynamic Modality Masking} We train \textsc{MAGE} with dynamic modality masking so that one conditional backbone supports the valid generation and extraction interfaces. For each training sample, we select one of three regimes: mixture-grounded target-source extraction with probability 0.70, mixture-free conditional generation with probability 0.20, and null-conditioning for classifier-free guidance with probability 0.10. In the extraction regime, the mixture latent is always provided together with at least one target-identifying condition. The valid extraction masks are mixture+text, mixture+visual, and mixture+text+visual. In the generation regime, the mixture branch is replaced by its learned null condition, and the valid masks are text-only, visual-only, and text+visual. We exclude mixture-only extraction because the mixture alone does not define a unique target source. This masking strategy exposes the same conditional flow model to all supported conditioning configurations without using task-specific output heads.

\partitle{Static-Frame Augmentation} When visual conditioning is present during training, we apply static-frame augmentation with probability 0.25 by repeating one randomly selected frame across the visual sequence. This exposes the visual-conditioning pathway to both sampled video-frame sequences and still-frame inputs. We use this only as a visual-conditioning augmentation; it is not intended to model fine-grained audio--visual synchronization.

\partitle{Optimization and Data Construction} We train the FluxFormer and trainable visual-conditioning layers with AdamW using learning rate $10^{-4}$, batch size 32, and gradient clipping at 5.0, while keeping the audio codec and CLAP text encoder frozen. Training runs for 2000 epochs. To increase variation in source pairings and temporal crops, the training set is repeated five times per epoch with independently sampled pairings and augmentations. Each extraction training example is constructed from two source clips. One clip is treated as the target source and the other as the non-target source. After audio loading and augmentation, the observed mixture is formed by combining the target and non-target waveforms. The target source, non-target source, and mixture are then encoded into the codec latent space. During audio loading, waveform amplitudes are randomly scaled by a factor sampled from $[0.5,1.5]$ and clipped to $[-1,1]$. Standard spatial augmentations are applied to sampled video frames during training, while evaluation uses fixed deterministic preprocessing.

\partitle{Inference} At inference time, we solve the learned latent flow with an Euler integrator using the same number of solver steps and the same guidance scale for all reported settings. For mixture-free conditional generation, the trajectory starts from Gaussian noise and is conditioned on text, visual input, or both, with the mixture branch set to its learned null condition. For mixture-grounded target-source extraction, the mixture latent is provided together with at least one target-identifying condition. The predicted target latent sequence is decoded to waveform using the frozen MixWavCodec decoder. The same latent representation, conditioning interface, decoder, and solver are used in both operating modes; only the selected condition set changes.

\partitle{Sampling-Step and Guidance Sensitivity} We evaluate sensitivity to the number of Euler solver steps and classifier-free guidance (CFG) scale for mixture-grounded target-source extraction. The trained model, conditioning setup, and evaluation procedure are kept fixed; only the solver-step count and CFG scale are varied. As shown in Table~\ref{tab:cfg_steps}, the default setting of 2 solver steps and CFG scale 1.5 gives the best SDR and SIR, indicating the strongest overall reconstruction and interference suppression. Increasing the guidance scale to 2.0 improves SAR but reduces SDR and SIR, suggesting a trade-off between artifact reduction and source selectivity.

\begin{table}[t]
\centering
\caption{
Sensitivity to Euler solver steps and classifier-free guidance scale for mixture-grounded target-source extraction.
All settings use the same trained model and evaluation protocol; only the number of solver steps and CFG scale are varied.
Higher values indicate better extraction quality, but SDR, SIR, and SAR reflect different reconstruction, interference, and artifact trade-offs.
}
\label{tab:cfg_steps}
\small
\setlength{\tabcolsep}{8pt}
\begin{tabular}{ccccc}
\toprule
\textbf{Steps} 
& \textbf{CFG} 
& \textbf{SDR} $\uparrow$ 
& \textbf{SIR} $\uparrow$ 
& \textbf{SAR} $\uparrow$ \\
\midrule
2 & 1.0 & 5.32 & 15.23 & 5.63 \\
2 & 1.5 & \textbf{6.67} & \textbf{17.99} & 7.88 \\
4 & 1.5 & 5.83 & 15.31 & 7.43 \\
4 & 2.0 & 5.73 & 14.64 & \textbf{16.12} \\
\bottomrule
\end{tabular}
\end{table}

\subsection{Evaluation Metrics}
\label{sec:evaluation_metrics} We evaluate \textsc{MAGE} under two separate protocols. In \emph{mixture-grounded target-source extraction}, the model receives an observed mixture and a target-identifying condition, and the goal is to recover the specified source. In \emph{mixture-free conditional generation}, the model receives text, visual input, or both, but no mixture audio. Since these settings measure different capabilities, we report source-separation metrics only for extraction and CLAP-based automatic alignment metrics for mixture-free generation.

\partitle{Source-Separation Metrics} For mixture-grounded target-source extraction, we report SDR, SIR, and SAR using the standard BSS-Eval protocol. Let $\hat{s}$ denote the predicted source and $s$ denote the ground-truth target source. SDR measures overall reconstruction quality, SIR measures suppression of non-target sources, and SAR measures artifact-related distortion. Higher values indicate better extraction quality. These metrics are used only in the mixture-grounded setting, where a reference target source is available. Dedicated separation methods are treated as task-specific baselines for extraction. \textsc{MAGE} is evaluated as a shared conditional model that supports both mixture-grounded extraction and mixture-free conditional generation, rather than as a replacement for specialized source-separation systems.

\partitle{Generation and Alignment Metrics} For mixture-free conditional generation, SDR, SIR, and SAR are not applicable because there is no observed mixture and no uniquely defined separated target. We therefore report CLAP-based automatic alignment metrics separately from the extraction metrics. CLAP text--audio similarity measures semantic alignment between the generated audio and the conditioning text prompt, while CLAP audio--audio similarity measures semantic similarity between generated audio and the reference target audio when a reference is available. Because CLAP is used both for text conditioning and automatic evaluation, these scores may favor semantic relationships captured by the CLAP embedding space. We therefore interpret CLAP-based metrics as automatic semantic-alignment indicators rather than substitutes for perceptual listening evaluation. These metrics are reported separately from SDR, SIR, and SAR to avoid conflating mixture-free generation quality with source-extraction quality.

\partitle{CLAP Audio--Audio Similarity}
We use CLAP audio--audio similarity as an automatic semantic-alignment measure between a predicted audio sample and a reference audio sample when such a reference is available. Let $f_a(\cdot)$ denote the CLAP audio encoder. Given predicted audio $\hat{s}$ and reference audio $s$, we compute
\begin{equation}
    \mathrm{Sim}_{\mathrm{AA}}
    =
    \frac{
        f_a(\hat{s})^\top f_a(s)
    }{
        \|f_a(\hat{s})\|_2 \, \|f_a(s)\|_2
    }.
\end{equation}
Higher values indicate stronger similarity in the CLAP audio embedding space. We use this metric to measure semantic correspondence to the reference target audio, while SDR, SIR, and SAR remain the signal-level metrics for mixture-grounded target-source extraction.

\partitle{CLAP Text--Audio Similarity}
We use CLAP text--audio similarity as an automatic measure of alignment between the predicted audio and the conditioning text prompt. Let $f_t(\cdot)$ denote the CLAP text encoder and $f_a(\cdot)$ denote the CLAP audio encoder. For prompt $c$ and predicted audio $\hat{s}$, we compute
\begin{equation}
    \mathrm{Sim}_{\mathrm{TA}}
    =
    \frac{
        f_t(c)^\top f_a(\hat{s})
    }{
        \|f_t(c)\|_2 \, \|f_a(\hat{s})\|_2
    }.
\end{equation}
Higher values indicate stronger text--audio alignment in the CLAP embedding space. Since CLAP is also used for text conditioning, we interpret CLAP-based scores as automatic semantic-alignment indicators rather than substitutes for perceptual listening evaluation. Together, the metrics are used under task-specific protocols. SDR, SIR, and SAR evaluate signal-level reconstruction, interference suppression, and artifact-related distortion for mixture-grounded target-source extraction. CLAP audio--audio similarity evaluates semantic similarity to the reference target audio when available, and CLAP text--audio similarity evaluates alignment with the conditioning text prompt. Since CLAP is also used for text conditioning, we interpret CLAP-based scores as automatic semantic-alignment indicators rather than substitutes for perceptual listening evaluation. We report these metrics separately to avoid conflating source-extraction quality with mixture-free generation quality.

\subsection{Fusion Operator Ablation}
\label{sec:fusion_ablation} We ablate how AVNA-aligned visual features are incorporated into the audio latent stream. All variants use the same backbone, training schedule, conditioning encoders, AVNA alignment, and evaluation protocol; they differ only in the visual fusion operator. We evaluate each variant on mixture-grounded target-source extraction and report SDR, SIR, and SAR in Table~\ref{tab:fusion}. We compare four alternatives against cross-gated modulation. Additive fusion directly adds projected visual features $W_s\bar{s}$ to the audio representation $h$. Gated residual fusion applies a learned gate to the projected visual features before adding them to the audio stream. Self-attention and cross-attention allow audio and visual tokens to interact through attention-based updates. These operators provide increasingly flexible feature interaction, but they either add visual information into the audio stream or update the audio representation through attention.

Cross-Gated Modulation instead uses the AVNA-aligned visual features to compute a multiplicative gate over the audio tokens:
\begin{equation}
    h_{\mathrm{out}}
    =
    h
    \odot
    \sigma(W_s\bar{s}+b_s).
\end{equation}
This design conditions the audio representation by modulating existing audio-token features rather than directly adding visual features. As shown in Table~\ref{tab:fusion}, cross-gated modulation obtains the best SDR, SIR, and SAR among the tested operators. The largest improvement is in SIR, suggesting that visual gating is particularly useful for suppressing non-target sources in the mixture-grounded extraction setting. This supports the use of aligned visual features as a conditioning gate rather than as directly fused auxiliary tokens.

\begin{table}[t]
\centering
\caption{
Fusion-operator ablation for mixture-grounded target-source extraction on MUSIC.
All variants use the same backbone, training setup, conditioning encoders, and AVNA-aligned visual features; they differ only in how visual information is incorporated into the audio latent stream.
Higher values indicate better extraction quality.
}
\label{tab:fusion}
\small
\setlength{\tabcolsep}{8pt}
\begin{tabular}{lccc}
\toprule
\textbf{Fusion Operator} 
& \textbf{SDR} $\uparrow$ 
& \textbf{SIR} $\uparrow$ 
& \textbf{SAR} $\uparrow$ \\
\midrule
Additive fusion 
& 2.09 & 3.53 & 4.32 \\
Gated residual  
& 2.41 & 4.61 & 5.37 \\
Self-attention  
& 2.42 & 4.64 & 5.45 \\
Cross-attention 
& 2.49 & 5.64 & 6.45 \\
\midrule
\textbf{Cross-gated modulation (ours)}
& \textbf{6.67} 
& \textbf{17.99} 
& \textbf{7.88} \\
\bottomrule  
\end{tabular}
\end{table}

\begin{figure}[t]
\centering
\includegraphics[width=\linewidth]{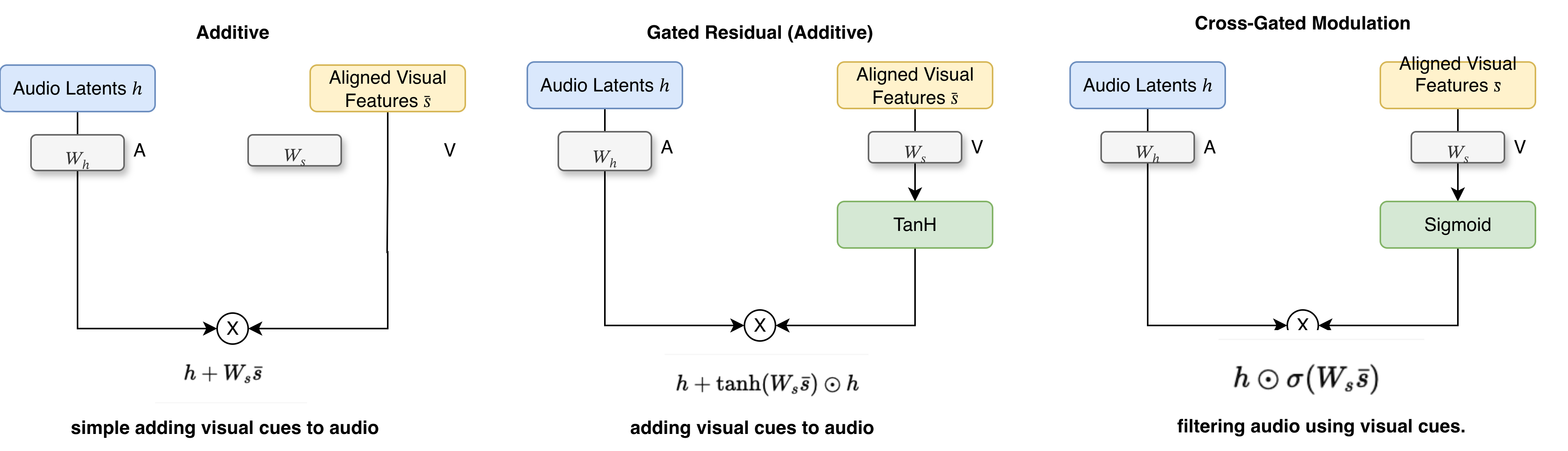}
\caption{
Illustration of representative visual fusion strategies.
\textbf{Left:} Additive fusion directly adds projected visual features $W_s\bar{s}$ to the audio representation $h$.
\textbf{Middle:} Gated residual fusion applies a learned gate to the projected visual features before adding them to the audio representation.
\textbf{Right:} Cross-gated modulation uses AVNA-aligned visual features to compute a gate, $h \odot \sigma(W_s\bar{s}+b_s)$, which modulates the audio representation.
Quantitative comparisons across all tested fusion operators are reported in Table~\ref{tab:fusion}.
}
\label{fig:cross-add}
\end{figure}

\subsection{Temporal vs. Static Visual Conditioning}
\label{sec:visual_temporal} We compare three visual-conditioning settings for mixture-grounded target-source extraction: static-image conditioning, shuffled-frame conditioning, and temporal-frame conditioning. In the static-image setting, one frame is sampled from the video segment and repeated across the visual sequence. In the shuffled-frame setting, the model receives the sampled frames but with their temporal order randomly permuted. In the temporal-frame setting, the model receives the uniformly sampled frame sequence in its original order. This comparison separates instrument- and scene-level evidence from temporal ordering effects. Table~\ref{tab:visual_temporal} shows that temporal-frame conditioning gives the best SDR and SIR. The improvement over static-image conditioning indicates that using multiple frames provides more useful target-identifying evidence than a single repeated frame. The improvement over shuffled-frame conditioning suggests that preserving the original frame order is beneficial for mixture-grounded extraction. The static-image variant obtains the highest SAR, suggesting fewer artifact-related distortions but weaker source selectivity. Overall, these results support the use of sampled temporal frame sequences, while also showing a trade-off between interference suppression and artifact control.

\begin{table}[t]
\centering
\caption{
Effect of visual temporal order on mixture-grounded target-source extraction on MUSIC.
Static-image conditioning repeats one sampled frame, shuffled-frame conditioning randomizes the sampled frame order, and temporal-frame conditioning preserves the original sampled frame sequence.
Higher values indicate better extraction quality.
}
\label{tab:visual_temporal}
\small
\setlength{\tabcolsep}{8pt}
\begin{tabular}{lccc}
\toprule
\textbf{Visual Conditioning} 
& \textbf{SDR} $\uparrow$ 
& \textbf{SIR} $\uparrow$ 
& \textbf{SAR} $\uparrow$ \\
\midrule
Static image 
& 4.12 
& 9.01 
& \textbf{8.73} \\
Shuffled frames 
& 5.12 
& 15.34 
& 5.34 \\
Temporal frames 
& \textbf{6.67} 
& \textbf{17.99} 
& 7.88 \\
\bottomrule
\end{tabular}
\end{table}


\vspace{-10pt}


{\small
\begin{thebibliography}{29}
\providecommand{\natexlab}[1]{#1}
\providecommand{\url}[1]{\texttt{#1}}
\expandafter\ifx\csname urlstyle\endcsname\relax
  \providecommand{\doi}[1]{doi: #1}\else
  \providecommand{\doi}{doi: \begingroup \urlstyle{rm}\Url}\fi

\bibitem[Chatterjee et~al.(2021)Chatterjee, Le~Roux, Ahuja, and
  Cherian]{chatterjee2021visual}
Moitreya Chatterjee, Jonathan Le~Roux, Narendra Ahuja, and Anoop Cherian.
\newblock Visual scene graphs for audio source separation.
\newblock In \emph{Proceedings of the IEEE/CVF International Conference on
  Computer Vision (ICCV)}, pages 1204--1213, 2021.

\bibitem[Chen et~al.(2023)Chen, Zhang, Lian, Yang, Zeng, and
  Shi]{chen2023iquery}
Jiaben Chen, Renrui Zhang, Dongze Lian, Jiaqi Yang, Ziyao Zeng, and Jianbo Shi.
\newblock iquery: Instruments as queries for audio-visual sound separation.
\newblock In \emph{Proceedings of the IEEE/CVF Conference on Computer Vision
  and Pattern Recognition (CVPR)}, pages 14675--14686, 2023.

\bibitem[Copet et~al.(2023)Copet, Kreuk, Gat, Remez, Kant, Synnaeve, Adi, and
  D{\'e}fossez]{copet2023simple}
Jade Copet, Felix Kreuk, Itai Gat, Tal Remez, David Kant, Gabriel Synnaeve,
  Yossi Adi, and Alexandre D{\'e}fossez.
\newblock Simple and controllable music generation.
\newblock In \emph{Advances in Neural Information Processing Systems}. Curran
  Associates, Inc., 2023.

\bibitem[D{\'e}fossez(2021)]{defossez2021hybrid}
Alexandre D{\'e}fossez.
\newblock Hybrid spectrogram and waveform source separation.
\newblock In \emph{Proceedings of the ISMIR 2021 Workshop on Music Source
  Separation}, 2021.

\bibitem[D{\'e}fossez et~al.(2019)D{\'e}fossez, Usunier, Bottou, and
  Bach]{defossez2019music}
Alexandre D{\'e}fossez, Nicolas Usunier, L{\'e}on Bottou, and Francis Bach.
\newblock Music source separation in the waveform domain.
\newblock \emph{arXiv preprint arXiv:1911.13254}, 2019.

\bibitem[Dong et~al.(2023)Dong, Takahashi, Mitsufuji, McAuley, and
  Berg-Kirkpatrick]{dong2023clipsep}
Hao-Wen Dong, Naoya Takahashi, Yuki Mitsufuji, Julian McAuley, and Taylor
  Berg-Kirkpatrick.
\newblock Clipsep: Learning text-queried sound separation with noisy unlabeled
  videos.
\newblock In \emph{Proceedings of International Conference on Learning
  Representations (ICLR)}, 2023.

\bibitem[Gan et~al.(2020)Gan, Huang, Zhao, Tenenbaum, and
  Torralba]{gan2020music}
Chuang Gan, Deng Huang, Hang Zhao, Joshua~B. Tenenbaum, and Antonio Torralba.
\newblock Music gesture for visual sound separation.
\newblock In \emph{Proceedings of the IEEE/CVF Conference on Computer Vision
  and Pattern Recognition (CVPR)}, pages 10478--10487, 2020.

\bibitem[Gao and Grauman(2019)]{gao2019co}
Ruohan Gao and Kristen Grauman.
\newblock Co-separating sounds of visual objects.
\newblock In \emph{Proceedings of the IEEE/CVF International Conference on
  Computer Vision (ICCV)}, pages 3879--3888, 2019.

\bibitem[Huang et~al.(2024)Huang, Liang, Tian, Kumar, and Xu]{huang2024accv}
C. Huang, S. Liang, Y. Tian, A. Kumar, and C. Xu.
\newblock High-quality visually-guided sound separation from diverse
  categories.
\newblock In \emph{Proceedings of the Asian Conference on Computer Vision
  (ACCV)}, pages 35--49, 2024.

\bibitem[Huang et~al.(2025)Huang, Liang, Tian, Kumar, and Xu]{huang2025davis}
Chao Huang, Susan Liang, Yapeng Tian, Anurag Kumar, and Chenliang Xu.
\newblock High-quality sound separation across diverse categories via
  visually-guided generative modeling.
\newblock \emph{arXiv preprint arXiv:2509.22063}, 2025.

\bibitem[Kong et~al.(2023)Kong, Chen, Liu, Du, Berg-Kirkpatrick, Dubnov, and
  Plumbley]{kong2023universal}
Qiuqiang Kong, Ke Chen, Haohe Liu, Xingjian Du, Taylor Berg-Kirkpatrick, Shlomo
  Dubnov, and Mark~D Plumbley.
\newblock Universal source separation with weakly labelled data.
\newblock \emph{arXiv preprint arXiv:2305.07447}, 2023.

\bibitem[Liu et~al.(2024)Liu, Yuan, Liu, Mei, Kong, Tian, Wang, Wang, Wang, and
  Plumbley]{liu2024audioldm2}
Haohe Liu, Yi Yuan, Xubo Liu, Xinhao Mei, Qiuqiang Kong, Qiao Tian, Yuping
  Wang, Wenwu Wang, Yuxuan Wang, and Mark~D. Plumbley.
\newblock {AudioLDM 2}: Learning holistic audio generation with self-supervised
  pretraining.
\newblock \emph{IEEE/ACM Transactions on Audio, Speech, and Language
  Processing}, 32:\penalty0 2871--2883, 2024.

\bibitem[Liu et~al.(2022)Liu, Liu, Kong, Mei, Zhao, Huang, Plumbley, and
  Wang]{AudioSep}
Xubo Liu, Haohe Liu, Qiuqiang Kong, Xinhao Mei, Jinzheng Zhao, Qiushi Huang,
  Mark~D Plumbley, and Wenwu Wang.
\newblock Separate what you describe: Language-queried audio source separation.
\newblock In \emph{Proc. Interspeech 2022}, pages 1801--1805, 2022.

\bibitem[Lu et~al.(2024)Lu, Wang, Kong, and Hung]{Lu2023MusicSS}
Wei-Tsung Lu, Ju-Chiang Wang, Qiuqiang Kong, and Yun-Ning Hung.
\newblock Music source separation with band-split rope transformer.
\newblock In \emph{ICASSP 2024-2024 IEEE International Conference on Acoustics,
  Speech and Signal Processing (ICASSP)}, pages 481--485. IEEE, 2024.

\bibitem[Prajwal et~al.(2024)Prajwal, Shi, Le, Vyas, Tjandra, Luthra, Guo,
  Wang, Afouras, Kant, et~al.]{prajwalmusicflow}
KR Prajwal, Bowen Shi, Matthew Le, Apoorv Vyas, Andros Tjandra, Mahi Luthra,
  Baishan Guo, Huiyu Wang, Triantafyllos Afouras, David Kant, et~al.
\newblock {MusicFlow}: Cascaded flow matching for text guided music generation.
\newblock In \emph{International Conference on Learning Representations}, 2024.

\bibitem[Radford et~al.(2021)Radford, Kim, Hallacy, Ramesh, Goh, Agarwal,
  Sastry, Askell, Mishkin, Clark, et~al.]{Radford2021Learning}
Alec Radford, Jong~Wook Kim, Chris Hallacy, Aditya Ramesh, Gabriel Goh,
  Sandhini Agarwal, Girish Sastry, Amanda Askell, Pamela Mishkin, Jack Clark,
  et~al.
\newblock Learning transferable visual models from natural language
  supervision.
\newblock In \emph{International conference on machine learning}, pages
  8748--8763. PmLR, 2021.

\bibitem[Rafii et~al.(2017)Rafii, Liutkus, St{\"o}ter, Mimilakis, and
  Bittner]{musdb18}
Zafar Rafii, Antoine Liutkus, Fabian-Robert St{\"o}ter, Stylianos~Ioannis
  Mimilakis, and Rachel Bittner.
\newblock The {MUSDB18} corpus for music separation, 2017.

\bibitem[Rouard et~al.(2023)Rouard, Massa, and D{\'e}fossez]{rouard2022hybrid}
Simon Rouard, Francisco Massa, and Alexandre D{\'e}fossez.
\newblock Hybrid transformers for music source separation.
\newblock In \emph{ICASSP 23}, 2023.

\bibitem[Shi et~al.(2025)Shi, Tjandra, Hoffman, Wang, Wu, Gao, Richter, Le,
  Vyas, Chen, Feichtenhofer, Doll{\'a}r, Hsu, and Lee]{shi2025samaudio}
B. Shi, A. Tjandra, J. Hoffman, H. Wang, Y.C. Wu, L. Gao, J. Richter, M. Le, A.
  Vyas, S. Chen, C. Feichtenhofer, P. Doll{\'a}r, W.N. Hsu, and A. Lee.
\newblock Sam audio: Segment anything in audio.
\newblock \emph{arXiv preprint arXiv:2512.18099}, 2025.

\bibitem[Spiertz and Gnann(2009)]{spiertz2009source}
M. Spiertz and V. Gnann.
\newblock Source-filter based clustering for monaural blind source separation.
\newblock In \emph{Proceedings of the 12th International Conference on Digital
  Audio Effects}, page~6, 2009.

\bibitem[Tian et~al.(2021)Tian, Hu, and Xu]{tian2021cyclic}
Yapeng Tian, Di Hu, and Chenliang Xu.
\newblock Cyclic co-learning of sounding object visual grounding and sound
  separation.
\newblock In \emph{Proceedings of the IEEE/CVF Conference on Computer Vision
  and Pattern Recognition (CVPR)}, pages 2745--2754, 2021.

\bibitem[Tian et~al.(2026)Tian, Liu, Jin, Yuan, Xue, Tan, Chen, Xue, and
  Guo]{tian2025audiox}
Zeyue Tian, Zhaoyang Liu, Yizhu Jin, Ruibin Yuan, Liumeng Xue, Xu Tan, Qifeng
  Chen, Wei Xue, and Yike Guo.
\newblock {AudioX}: A unified framework for anything-to-audio generation.
\newblock In \emph{International Conference on Learning Representations}, 2026.

\bibitem[Wang et~al.(2025)Wang, Hai, Lu, Thakkar, Elhilali, and
  Dehak]{wang2025soloaudio}
Helin Wang, Jiarui Hai, Yen-Ju Lu, Karan Thakkar, Mounya Elhilali, and Najim
  Dehak.
\newblock Soloaudio: Target sound extraction with language-oriented audio
  diffusion transformer.
\newblock In \emph{ICASSP 2025-2025 IEEE International Conference on Acoustics,
  Speech and Signal Processing (ICASSP)}, pages 1--5. IEEE, 2025.

\bibitem[Wen et~al.(2025)Wen, Chen, Seetharaman, Nieto, Su, Kumar, Kim,
  Smaragdis, Jin, and Salamon]{wen2025promptsep}
Yutong Wen, Ke Chen, Prem Seetharaman, Oriol Nieto, Jiaqi Su, Rithesh Kumar,
  Minje Kim, Paris Smaragdis, Zeyu Jin, and Justin Salamon.
\newblock Promptsep: Generative audio separation via multimodal prompting.
\newblock \emph{arXiv preprint arXiv:2511.04623}, 2025.

\bibitem[Wu et~al.(2023)Wu, Chen, Zhang, Hui, Berg-Kirkpatrick, and
  Dubnov]{wu2023clap}
Yusong Wu, Ke Chen, Tianyu Zhang, Yuchen Hui, Taylor Berg-Kirkpatrick, and
  Shlomo Dubnov.
\newblock Large-scale contrastive language-audio pretraining with feature
  fusion and keyword-to-caption augmentation.
\newblock In \emph{ICASSP 2023-2023 IEEE International Conference on Acoustics,
  Speech and Signal Processing (ICASSP)}, pages 1--5. IEEE, 2023.

\bibitem[Yuan et~al.(2025)Yuan, Liu, Liu, Plumbley, and Wang]{flowsep}
Yi Yuan, Xubo Liu, Haohe Liu, Mark~D Plumbley, and Wenwu Wang.
\newblock Flowsep: Language-queried sound separation with rectified flow
  matching.
\newblock In \emph{ICASSP 2025-2025 IEEE International Conference on Acoustics,
  Speech and Signal Processing (ICASSP)}, pages 1--5. IEEE, 2025.

\bibitem[Zhao et~al.(2018)Zhao, Gan, Rouditchenko, Vondrick, McDermott, and
  Torralba]{zhao2018sound}
Hang Zhao, Chuang Gan, Andrew Rouditchenko, Carl Vondrick, Josh McDermott, and
  Antonio Torralba.
\newblock The sound of pixels.
\newblock In \emph{Proceedings of the European conference on computer vision
  (ECCV)}, pages 570--586, 2018.

\bibitem[Zhu and Rahtu(2022)]{zhu2022visually}
Lifei Zhu and Esa Rahtu.
\newblock Visually guided sound source separation and localization using
  self-supervised motion representations.
\newblock In \emph{Proceedings of the IEEE/CVF Winter Conference on
  Applications of Computer Vision (WACV)}, pages 1289--1299, 2022.

\bibitem[Zuo et~al.(2025)Zuo, You, Wu, Ren, Chen, Zhou, Lu, and
  Sun]{zuo2025gvmgen}
Heda Zuo, Weitao You, Junxian Wu, Shihong Ren, Pei Chen, Mingxu Zhou, Yujia Lu,
  and Lingyun Sun.
\newblock {GVMGen}: A general video-to-music generation model with hierarchical
  attentions.
\newblock In \emph{Proceedings of the AAAI Conference on Artificial
  Intelligence}, pages 23099--23107, 2025.

\end{thebibliography}

}
\end{document}